\documentclass[10pt,conference,letterpaper]{IEEEtran}
\IEEEoverridecommandlockouts

\usepackage{cite}
\usepackage{amsmath,amssymb,amsfonts}
\usepackage{graphicx}
\usepackage{textcomp}
\usepackage{xcolor}
\usepackage{booktabs}
\usepackage{multirow}
\usepackage{tabularx}
\usepackage{hyperref}

\def\BibTeX{{\rm B\kern-.05em{\sc i\kern-.025em b}\kern-.08em
    T\kern-.1667em\lower.7ex\hbox{E}\kern-.125emX}}

\begin{document}

\title{A Protocol-Guided LLM Agent for Quantum Program Synthesis and Execution}

\author{
\IEEEauthorblockN{
    Ming-Kang Ho\IEEEauthorrefmark{1}
    and Tai-Yue Li\IEEEauthorrefmark{1}
}
\IEEEauthorblockA{\IEEEauthorrefmark{1} National Center for High-performance Computing, National Institutes of Applied Research, Hsinchu, Taiwan}
\IEEEauthorblockA{
Emails: 2603013@niar.org.tw, tim312508@gmail.com}
}

\maketitle

\begin{abstract}
This study evaluates a protocol-guided large language model (LLM) agent workflow for quantum program synthesis and execution. A versioned YAML protocol specifies interface and quantum-semantic requirements while leaving circuit design to the model. The workflow combines Qiskit circuit generation, evaluator-guided repair, and quantum processing unit (QPU) deployment. A matched ablation evaluates four LLM endpoints using the variational quantum eigensolver (VQE) for H$_2$, the quantum approximate optimization algorithm (QAOA) for MaxCut, and Fashion-MNIST quantum-kernel classification. Across 240 trials, protocol guidance increased evaluator completion from 80.0\% to 99.2\%, reduced mean repairs from 2.56 to 0.75, and reduced mean agent time from 55.70 to 30.58~s. Provider-reported token use decreased for three of four endpoints. Completed VQE trials meeting chemical accuracy increased from 16/40 to 29/40, while valid QAOA approximation ratios remained comparable. Semantic review identified rank-one fidelity kernels, reinforcing the distinction between evaluator completion and semantic validity. A 210-run IBM QPU study evaluated the combined generation-and-selection pipelines and confirmed deployment feasibility. The results support protocol guidance as a software-reliability layer that must be complemented by quantum-semantic validation.
\end{abstract}


\section{Introduction}
Large language model (LLM) agents translate natural-language requests into executable actions. In quantum programming, however, API-compliant code may still use an unsuitable reference state, a commuting ``mixer,'' an incomplete cost Hamiltonian, or a global-phase-only feature map. Evaluation must therefore distinguish evaluator completion, quantum-semantic validity, and task quality.

ReAct uses environmental feedback to guide actions and repairs \cite{yao2023react}. Qiskit HumanEval, QuanBench, and its multi-framework extension expose functional and quantum-semantic failures in generated quantum code \cite{vishwakarma2024qiskithumaneval,guo2025quanbench,slim2026quanbenchplus}; application benchmarks assess circuit performance separately \cite{lubinski2023application}. Specification-based approaches such as QbC construct quantum programs from formal pre- and postconditions, providing a complementary route to correctness \cite{peduri2025qbc}. These studies motivate feedback-driven repair and layered evaluation but do not isolate a versioned domain protocol in a matched agent workflow.

This study asks whether a compact protocol improves reliability and efficiency without supplying a solution. PG and PF share the task request, base instructions, endpoint, evaluator, repair budget, and decoding settings; only PG receives the YAML package. It excludes completed circuits, gate sequences, optimized parameters, and expected answers. Both arms independently construct VQE \cite{peruzzo2014vqe}, QAOA \cite{farhi2014qaoa}, and quantum support vector machine (QSVM) circuits \cite{havlicek2019quantum}.

The contributions are an auditable generation-to-QPU workflow, a controlled package-level ablation across four endpoints and three tasks, and separate analyses of completion, semantics, task quality, and deployment. The heterogeneous endpoints test cross-family consistency rather than rank models.

\section{Methodology}

\begin{figure*}[!t]
    \centering
    \includegraphics[width=0.98\textwidth]{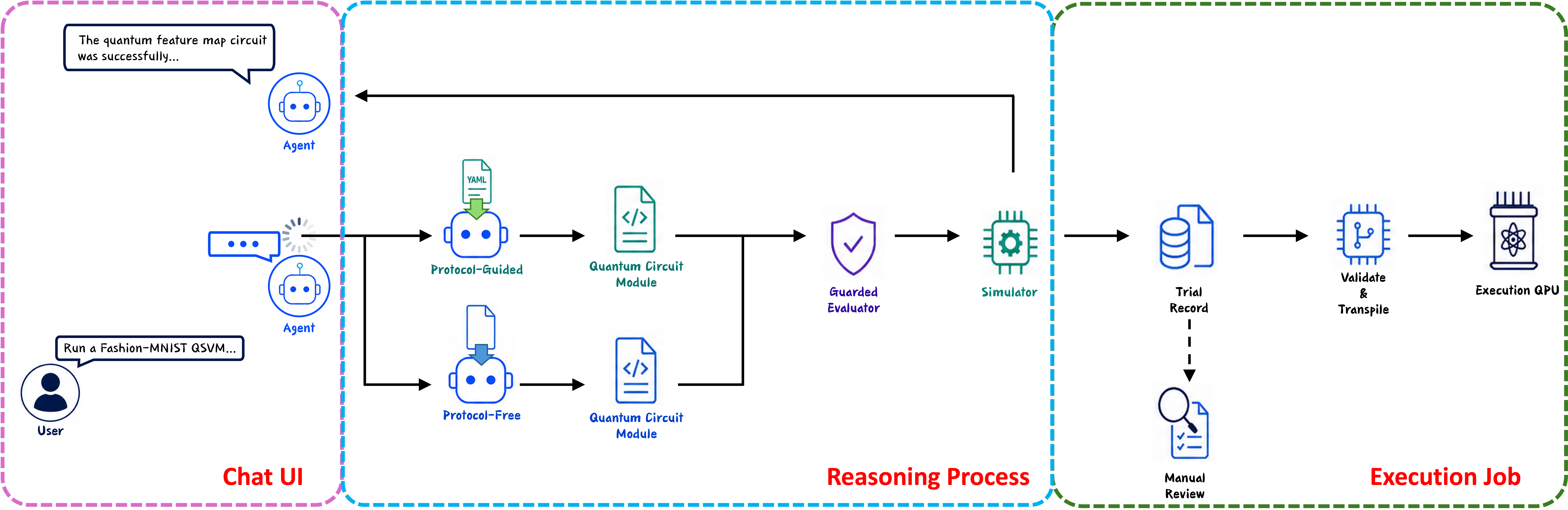}
    \caption{End-to-end workflow. Matched PG and PF trials share generation,
    evaluation, simulation, and feedback; only PG receives YAML guidance.
    Accepted records undergo semantic audit, and arm-specific selectors nominate
    candidates for ISA transpilation and QPU execution.}
    \label{fig:quantum-hub-workflow}
\end{figure*}

\subsection{Agent Workflow and Evaluation}
Figure~\ref{fig:quantum-hub-workflow} summarizes request handling, matched PG/PF execution, semantic review, and deployment. Both arms use the same OpenAI Agents SDK orchestration and generate the same artifact: a self-contained Python module that constructs a Qiskit circuit \cite{javadiabhari2024qiskit}. The module exposes required metadata and defines a no-argument \texttt{build\_circuit()} function returning a parameterized \texttt{QuantumCircuit}. The agent chooses the circuit structure, parameterization, and task metadata. A trusted harness implements the Hamiltonian or objective, preprocessing, classical optimization, simulation, and metrics from those declarations.

The common input comprises the task request and a generic contract requiring \texttt{APPLICATION}, \texttt{PROBLEM}, and \texttt{SEED}, the circuit-building function, and applicable optimizer and shot settings. It neither enumerates task-specific field values nor prescribes circuit design. Both arms also share the endpoint profile, tool schema, evaluator, diagnostics, decoding settings, and five-submission budget. The evaluator checks syntax, restricted calls, metadata, return type, qubit and operation limits, absence of mid-circuit measurements, and parameter presence. QAOA additionally requires \texttt{gamma} and \texttt{beta} roles; QSVM requires one data parameter per feature. Accepted candidates enter a fixed task evaluator, whereas rejections return structured errors or a bounded runtime trace for repair. Source, diagnostics, circuit evidence, latency, repairs, and provider-reported tokens are archived.

\subsection{Protocol Treatment and Semantic Audit}
The treatment is the complete YAML package appended only to PG. Relative to the common contract, it enumerates permitted \texttt{APPLICATION} values, task-specific \texttt{PROBLEM} fields, and applicable execution fields. It also adds general circuit semantics, task obligations, and non-solution guarantees. PF receives none of these clauses and must infer task-specific values from the request and common contract. The arm contrast therefore estimates the joint effect of interface disambiguation and semantic guidance rather than either component alone. The package uses schema \texttt{quantum-circuit-generation/v3} (SHA-256 prefix \texttt{307c9bca8446}); every PG manifest records the full hash. Its clauses require meaningful parameter use, an occupation reference and correlation-capable VQE ansatz, complete QAOA edge evolution and a noncommuting all-qubit mixer, and a reusable data-dependent entangling QSVM map. It supplies no completed circuit, gate order, depth, parameter value, optimized angle, label, or expected result.

Structural acceptance does not establish task fidelity. All 65 evaluator-complete QSVM records were screened by numerical kernel rank using tolerance $10^{-9}$; source review then examined the Qwen PG source repeated across ten rank-one records and one configuration-drift source. Ten PF Qwen QAOA records containing spaced multi-register count keys were reparsed after removing register separators. These post-hoc adjustments changed only audited task-quality summaries: no program was regenerated, and workflow metrics retain the original evaluator outcomes. One QSVM trial used a 40/20 rather than 20/10 split because the evaluator read sample counts from generated \texttt{PROBLEM} metadata. It remains in workflow summaries but is excluded from the PF GPT-OSS accuracy mean. The targeted audit may miss other semantic failures and is not an automated correctness proof.

\subsection{Candidate Selection and Deployment}
QPU deployment uses a separate selection stage. For each model--arm pair, a selector reviews ten H$_2$ records and nominates three candidates using numerical quality, quantum meaning, circuit structure, optimizer behavior, reliability, and hardware suitability; no fixed score is used. Only the PG selector receives additional semantic clauses. A deterministic validator checks assessment coverage, distinct identifiers, local success, and source evidence. Selected sources are hash-verified, re-executed locally, bound to optimized parameters, and transpiled to the target backend's ISA. Because generation and selection are both arm-specific, QPU contrasts evaluate the combined generation-and-selection pipelines rather than an isolated generation effect.

\begin{figure*}[!t]
    \centering
    \includegraphics[width=0.98\textwidth]{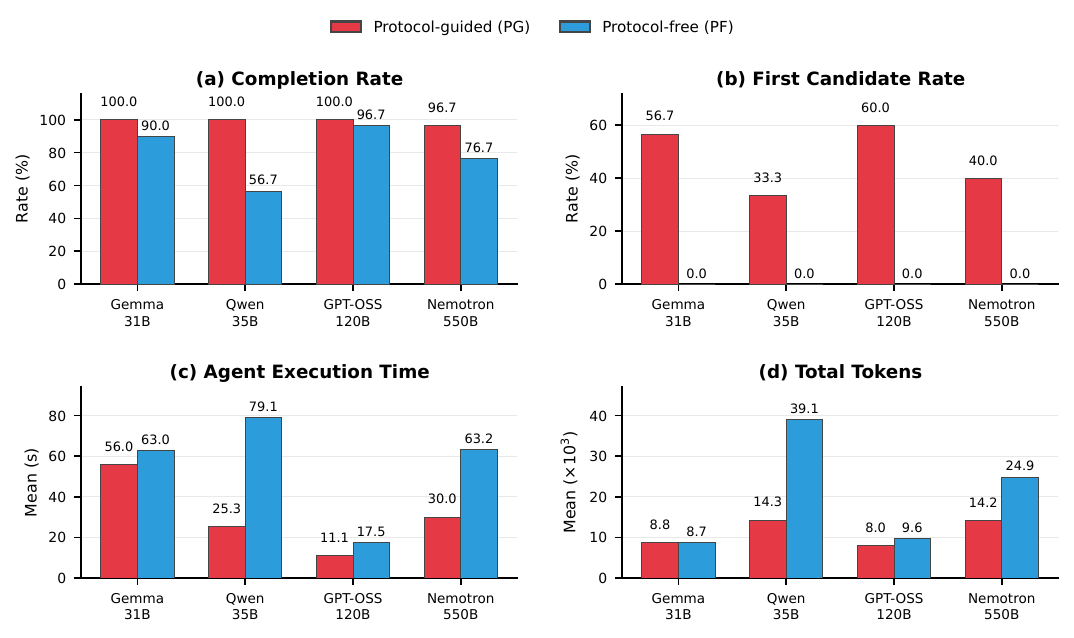}
    \caption{Workflow results by model (30 trials per arm): evaluator and
    first-candidate completion, agent time, and repairs. Cost means exclude
    missing telemetry; lower is better for time and repairs.}
    \label{fig:cross-model-workflow}
\end{figure*}

\section{Experimental Setup}
\noindent\textit{Paired benchmark design.}
The benchmark includes Gemma-4-31B-IT, Qwen3.6-35B-A3B, GPT-OSS-120B, and Nemotron-3-Ultra-550B-A55B. Qwen uses a local OpenAI-compatible endpoint; the others use NVIDIA-hosted endpoints. Within every model--task pair, both arms share the endpoint profile, task request, temperature-zero setting, evaluator, and five-submission budget; arm order alternates across repetitions. Ten repetitions per cell of the $4\times3\times2$ design yield 240 trials. The archived artifacts include prompts, endpoint settings, seeds, requirement and protocol hashes, generated sources, evaluator outputs, and repair histories.

\noindent\textit{Tasks.}
The tasks probe distinct semantic obligations and use seed 42. Two-qubit H$_2$ VQE tests reference-state and ansatz design at 0.735~\AA with a fixed Hamiltonian and exact ground energy $-1.8572750302$~Ha; energies are computed from statevectors. MaxCut tests graph coverage and mixer semantics using $p=1$ QAOA on an unweighted eight-node ring. Optimization uses exact probabilities, whereas final expected cuts and ratios use 1024 seeded samples against an optimum of eight. Both variational tasks use COBYLA with \texttt{maxiter}=100. Fashion-MNIST QSVM distinguishes Coat from Shirt and tests whether inputs induce meaningful state changes \cite{xiao2017fashion}. Standardization and four-component PCA are fitted only to 20 balanced training samples; angle scaling supplies four inputs to a generated four-qubit feature map. Its exact fidelity kernel feeds a fixed precomputed-kernel SVC evaluated on 10 test samples. This compact task probes feature-map semantics rather than classification at scale \cite{chen2025qsvm,sam2026dual}.

\noindent\textit{Metrics.}
Workflow metrics are evaluator completion, first-candidate completion, repairs, agent time, and tokens. Task metrics are VQE chemical accuracy ($|\Delta E|\leq1.6\times10^{-3}$~Ha), corrected QAOA approximation ratio, and semantically reviewed QSVM accuracy. Completion denotes contract satisfaction, not scientific validity. Wilson intervals, exact McNemar tests, and two-sided Wilcoxon signed-rank tests summarize within-benchmark consistency. Four PF failures lack complete cost telemetry, leaving 116 paired repair and time observations. Tokens are compared only within models because endpoint tokenizers differ; repetitions measure observed workflow robustness rather than population-level variation.

\noindent\textit{Hardware follow-up.}
An independent set of 80 H$_2$ trials ($4\times2\times10$), not pooled with the main benchmark, supports hardware selection. Eight arm-specific selectors nominate 24 candidates for \texttt{ibm\_fez}, \texttt{ibm\_marrakesh}, and \texttt{ibm\_kingston}. Each candidate runs three times with 1024 shots per measurement-basis circuit. Kingston provides the complete, closely timed subset; Fez and Marrakesh each lack one PF candidate. Nonrandom candidate selection and unsynchronized calibration limit these combined-pipeline comparisons to exploratory deployment evidence.

\begin{table*}[t]
\caption{Task quality by model and arm: VQE chemical passes out of ten,
corrected mean QAOA ratios, and reviewed QSVM accuracy (valid $n$).
A dash indicates no valid result.}
\label{tab:quality}
\centering
\footnotesize
\begin{tabular}{lcccccc}
\toprule
& \multicolumn{2}{c}{H$_2$ pass count} & \multicolumn{2}{c}{QAOA ratio} & \multicolumn{2}{c}{QSVM accuracy ($n$)} \\
\cmidrule(lr){2-3}\cmidrule(lr){4-5}\cmidrule(lr){6-7}
Model & PG & PF & PG & PF & PG & PF \\
\midrule
Gemma 31B & 9/10 & 3/10 & 0.7468 & 0.7469 & 0.570 (10) & 0.590 (10) \\
Qwen 35B & 10/10 & 7/10 & 0.7471 & 0.7468 & Invalid (rank 1) & -- \\
GPT-OSS 120B & 4/10 & 2/10 & 0.7469 & 0.7469 & 0.410 (10) & 0.522 (9) \\
Nemotron 550B & 6/10 & 4/10 & 0.7469 & 0.7469 & 0.510 (10) & 0.460 (5) \\
\bottomrule
\end{tabular}
\end{table*}

\section{Results}
Protocol guidance improved workflow reliability across endpoints, while task-quality effects varied. Figure~\ref{fig:cross-model-workflow} shows completion gains of 3.3--43.3 points. Overall, PG completed 119/120 trials (99.2\%; 95\% CI: 95.4--99.9\%) versus PF's 96/120 (80.0\%; 72.0--86.2\%); matched discordances favored PG 24 to one (McNemar $p=1.55\times10^{-6}$).

First-candidate completion was 33.3\%--60.0\% for PG and 0\% for PF. Of 120 PF first submissions, 87 used an unsupported \texttt{APPLICATION} and 81 a non-dictionary \texttt{PROBLEM}, with overlap; the gap therefore mainly reflects interface disambiguation, not semantic clauses alone. Mean repairs fell from 2.56 to 0.75 and cell-mean time from 55.70 to 30.58~s. Across 116 pairs, median differences were $-2$ repairs (Wilcoxon $p=1.14\times10^{-17}$) and $-11.11$~s ($p=7.40\times10^{-10}$). Tokens fell 16.8--63.5\% for three endpoints and rose 0.5\% for Gemma.

Table~\ref{tab:quality} shows the largest task gain for VQE: chemical passes increased from 16/40 to 29/40, with paired discordances favoring PG 16 to three ($p=0.0044$). Corrected QAOA ratios were nearly identical (0.7468--0.7471), making QAOA primarily a reliability and semantic-coverage probe.

QSVM marks the boundary of this benefit: PG accuracy was lower for Gemma and GPT-OSS, higher for Nemotron, and unavailable for Qwen comparison. All ten Qwen PG records reused one source that passed but produced rank-one kernels. The corrected PF Qwen QAOA aggregate covers all ten affected records; the PF GPT-OSS QSVM mean excludes one 40/20 split trial, retained in workflow summaries.

Across three IBM backends, 70 candidate--backend cells yielded 210 repetitions; all 24 Kingston candidates ran. Mean absolute error was 0.02420~Ha for PG and 0.01914~Ha for PF, while within-candidate deviation was 17.2\% lower for PG (0.00569 versus 0.00687~Ha). These combined-pipeline results demonstrate deployment and repeatability, not a QPU accuracy advantage.

\section{Discussion}
Workflow gains should not be interpreted as a general improvement in quantum reasoning. Most PF first submissions failed on \texttt{APPLICATION} or \texttt{PROBLEM} metadata, whereas corrected QAOA quality changed little once implementations were valid. The package thus primarily reduces request-to-artifact ambiguity. VQE gains suggest an additional benefit from task-specific semantics, but interface and semantic contributions cannot be isolated because their clauses were applied jointly.

The QSVM case exposes a complementary failure: the repeated Qwen feature map satisfied structural requirements and used all inputs, yet produced rank-one fidelity kernels. Structural checks should therefore be paired with behavioral tests of input sensitivity, kernel rank, cost coverage, mixer noncommutativity, and reference-state preparation, moving semantic assessment into the trusted harness.

Hardware interpretation is also bounded. Because generation and arm-specific selection differed, the Kingston repeatability cannot be attributed to generation alone. A common selector, matched bound circuits, synchronized execution windows, and calibration snapshots would enable a stronger comparison.

\section{Conclusion}
The complete package raised completion from 80.0\% to 99.2\%, reduced mean repairs from 2.56 to 0.75, and reduced mean agent time from 55.70 to 30.58~s without supplying circuits or answers. Because the package also disambiguates task fields, these gains combine interface and semantic guidance. Task-quality effects were application-dependent: chemically accurate VQE completions increased from 16/40 to 29/40, corrected QAOA ratios were nearly unchanged, and rank-one QSVM kernels confirmed that contract compliance does not replace semantic validation.

The 210 QPU repetitions establish an auditable combined-pipeline deployment path rather than an accuracy advantage. Protocol guidance should therefore be paired with automated semantic checks. Future work should separate interface and semantic clauses, diversify instances, and compare matched circuits under synchronized calibration.

\section*{Acknowledgment}
The authors acknowledge NCHC for computational support; IBM Quantum, IQM Resonance, and Amazon Braket for quantum resources; and NVIDIA for LLM resources. ChatGPT assisted with language editing; the authors reviewed all revisions.

\bibliographystyle{ieeetr}
\bibliography{reference_ty}

\end{document}